\documentclass[prd,amsmath,amssymb,superscriptaddress,preprintnumbers,twocolumn,nofootinbib,10pt]{revtex4-1}
\pdfoutput=1
\usepackage{graphicx}
\usepackage{dcolumn}
\usepackage{bm}
\usepackage{amssymb}
\usepackage{latexsym}
\usepackage{booktabs}
\usepackage{amsmath}
\usepackage{multirow}
\usepackage{url}
\usepackage{footnote}
\usepackage{float}
\usepackage{threeparttable}
\usepackage[colorlinks=true, linkcolor=blue, citecolor=blue]{hyperref}
\usepackage[bottom]{footmisc}

\usepackage[normalem]{ulem}
\usepackage{color}
\usepackage{array}
\usepackage{enumerate}
\usepackage{adjustbox}

\usepackage{makecell}
\usepackage{diagbox}
\usepackage{epstopdf}
\usepackage{epsfig}
\usepackage{longtable}
\usepackage{supertabular}
\usepackage{algorithm}
\usepackage{pifont}
\usepackage{algorithmic}
\usepackage{changepage}
\usepackage{setspace}
\begin{document}

\title{Constraints on interacting dark energy with a constant equation of state in light of DESI DR2: The role of the interaction term} 

\author{Zhao-Yu Li}
\affiliation{Liaoning Key Laboratory of Cosmology and Astrophysics, College of Sciences, Northeastern University, Shenyang 110819, China}

\author{Tian-Nuo Li}\thanks{Corresponding author}\email{litiannuo@ucas.ac.cn}
\affiliation{School of Fundamental Physics and Mathematical Sciences, Hangzhou Institute for Advanced Study, UCAS, Hangzhou 310024, China}

\author{Guo-Hong Du}
\affiliation{Liaoning Key Laboratory of Cosmology and Astrophysics, College of Sciences, Northeastern University, Shenyang 110819, China}

\author{Sheng-Han Zhou}
\affiliation{Liaoning Key Laboratory of Cosmology and Astrophysics, College of Sciences, Northeastern University, Shenyang 110819, China}

\author{Yun-He Li}
\affiliation{Liaoning Key Laboratory of Cosmology and Astrophysics, College of Sciences, Northeastern University, Shenyang 110819, China}

\author{Jing-Fei Zhang}
\affiliation{Liaoning Key Laboratory of Cosmology and Astrophysics, College of Sciences, Northeastern University, Shenyang 110819, China}

\author{Xin Zhang}\thanks{Corresponding author}\email{zhangxin@mail.neu.edu.cn}
\affiliation{Liaoning Key Laboratory of Cosmology and Astrophysics, College of Sciences, Northeastern University, Shenyang 110819, China}
\affiliation{MOE Key Laboratory of Data Analytics and Optimization for Smart Industry, Northeastern University, Shenyang 110819, China}
\affiliation{National Frontiers Science Center for Industrial Intelligence and Systems Optimization, Northeastern University, Shenyang 110819, China}

\begin{abstract}
Recent DESI baryon acoustic oscillation (BAO) analyses suggest a preference for an evolving dark energy equation of state (EoS) that crosses the phantom divide $w=-1$. Interactions between dark energy and dark matter can also produce an apparent phantom-divide crossing in the effective dark energy EoS. In this work, we investigate interacting dark energy (IDE) models with a constant EoS parameter $w$, using cosmic microwave background (CMB) observations from Planck, ACT, and SPT, DESI DR2 BAO measurements, and type Ia supernova data from PantheonPlus, DES-Dovekie, and DESY5. We consider four typical forms of the interaction term $Q$. Our results show that the IDE models with $Q\propto\rho_{\rm de}$ generally exhibit a stronger preference for nonzero dark-sector interactions. In particular, for the CMB+DESI+DES-Dovekie data, the IDE model with $Q=\beta H_0\rho_{\rm de}$ yields $\beta=-2.52^{+0.80}_{-0.54}$ and $w=-1.510^{+0.170}_{-0.120}$, corresponding to deviations from $\beta=0$ and $w=-1$ at the $3.2\sigma$ and $3.0\sigma$ levels, respectively. Interestingly, the effective EoS of the IDE model with $Q=\beta H_0\rho_{\rm de}$ exhibits a redshift evolution resembling that of the $w_0w_a$CDM model, crossing the phantom divide. Bayesian evidence further indicates that this IDE model is favored over $\Lambda$CDM and the other three IDE models. Our analysis helps identify which interaction forms better account for current cosmological observations.
\end{abstract}

\maketitle
\section{Introduction}
\label{sec1}

Over the past few decades, the standard $\Lambda$CDM model has successfully described a broad range of cosmological observations, including the cosmic microwave background (CMB), baryon acoustic oscillations (BAO), and type Ia supernova (SN) observations. Nevertheless, this model still faces several theoretical and observational challenges. On the theoretical side, the cosmological constant $\Lambda$ suffers from the ``fine-tuning'' and ``cosmic coincidence'' problems~\cite{Sahni:1999gb,Bean:2005ru}, while on the observational side, the persistent $H_0$ tension between early-universe CMB inferences and late-time distance-ladder measurements may indicate that the standard cosmological framework remains incomplete~\cite{Planck:2018vyg,Riess:2019qba,Riess:2021jrx,Verde:2019ivm,Guo:2018ans,DiValentino:2019ffd,Vagnozzi:2019ezj,DiValentino:2020zio,DiValentino:2021izs,Vagnozzi:2021gjh,Abdalla:2022yfr,Schoneberg:2021qvd,Giare:2023xoc,Gao:2021xnk,Hu:2023jqc,Poulin:2023lkg,Vagnozzi:2023nrq,Cai:2026swf,Schoneberg:2026vaf,Sabogal:2026ipu,Pedrotti:2026dwj,Pedrotti:2025ccw,Du:2026qtq,Song:2025ddm,Liu:2025evk,Liu:2024gne}. More recently, BAO measurements from the Dark Energy Spectroscopic Instrument (DESI) Data Release~2 (DR2), when combined with CMB data and different SN samples, show a preference for dynamical dark energy within the $w_0w_a$CDM framework, reaching up to $4.2\sigma$~\cite{DESI:2025zgx}. The preferred evolution exhibits a quintom-like behavior, with the dark energy EoS transitioning from $w<-1$ in the past to $w>-1$ at present~\cite{Feng:2004ad,Zhang:2005kj}. This result has further stimulated extensive investigations of dynamical dark energy and other possible extensions of the $\Lambda$CDM model~\cite{Cortes:2024lgw,Shlivko:2024llw,Luongo:2024fww,Gialamas:2024lyw,Wang:2024dka,Ye:2024ywg,Tada:2024znt,DESI:2024kob,Bhattacharya:2024hep,Ramadan:2024kmn,Malekjani:2024bgi,Giare:2024gpk,Reboucas:2024smm,Park:2024pew,Li:2024qus,Jiang:2024xnu,Yang:2024kdo,Li:2025vuh,Wolf:2025jed,Shajib:2025tpd,Chaussidon:2025npr,Pang:2025lvh,RoyChoudhury:2025dhe,Paliathanasis:2025cuc,Scherer:2025esj,Giare:2024oil,Liu:2025mub,Teixeira:2025czm,Specogna:2025guo,Cheng:2025lod,Herold:2025hkb,Ozulker:2025ehg,Ormondroyd:2025iaf,Silva:2025twg,Ishak:2025cay,Fazzari:2025lzd,Zhang:2025lam,Cheng:2025hug,Song:2025bio,Li:2025ops,Lee:2025pzo,Wu:2025vfs,Santos:2025wiv,Khoury:2025txd,Li:2025ula,Kessler:2025kju,Smith:2025icl,Li:2025owk,Li:2025dwz,Li:2026hwq,Li:2026xaz,Giare:2024smz,Lemos:2025bgy,You:2025uon,Qiu:2025oop,Ren:2026jyw,Turner:2026vzp,Ribeiro:2026lyd,Bella:2026zuk,Wang:2026wrk,Wang:2025znm,Yang:2025gaz,Wang:2026vqw,Yang:2026qmz,Pan:2025qwy,Yang:2026yaq,Du:2025xes,Pedrotti:2026ncd,Yao:2025twv,Mukherjee:2026lnb,Yang:2026vti,Fortunato:2025qxc,Kulhkampa:2026bfg,Xu:2026sbw,Zhang:2026axl,Ling:2025lmw,Jiang:2026ngo}. For recent reviews, see, e.g., Refs.~\cite{CosmoVerseNetwork:2025alb,Giare:2025pzu,Cai:2025mas,Anchordoqui:2026hys,Li:2026asg,Giare:2026oti}.

However, the preference for dynamical dark energy indicated by DESI DR2 does not uniquely imply that the dark energy equation of state (EoS) itself evolves with time. Distance probes such as BAO and SN primarily constrain the cosmic expansion history and the distance--redshift relation rather than directly measuring the dark energy EoS $w(z)$, so similar observational behavior may arise from different physical mechanisms. Interpreting these deviations in terms of an evolving $w(z)$ generally assumes that dark matter and dark energy are separately conserved. If this assumption is relaxed, allowing energy and momentum exchange between the two dark components modifies the evolution of the dark matter and dark energy densities, $\rho_{\rm c}(z)$ and $\rho_{\rm de}(z)$, and consequently the late-time cosmic expansion.

Interacting dark energy (IDE) therefore provides a physically distinct possibility for interpreting the departures from $\Lambda$CDM suggested by current observations~\cite{Zhang:2005rj,Zhang:2005rg,Zhang:2007uh,Salvatelli:2013wra,Li:2015vla,Guo:2018gyo,Murgia:2016ccp,Kumar:2017dnp,DiValentino:2017iww,DiValentino:2020vnx,Pan:2023mie,Forconi:2023hsj, Pourtsidou:2016ico,Nunes:2021zzi,Wang:2021kxc,Lucca:2020zjb,Zhai:2023yny,Becker:2020hzj,Hoerning:2023hks,Giare:2024ytc,Escamilla:2023shf,Li:2024qso,Halder:2024uao, Castello:2023zjr,Yao:2022kub,Li:2023gtu,Mishra:2023ueo,Nunes:2016dlj,Silva:2025hxw,vanderWesthuizen:2025rip,Zhang:2025dwu,Li:2025muv,Li:2025owk,Lyu:2025nsd,Yang:2025uyv,Pan:2025qwy,Wang:2016lxa,Wang:2024vmw,Feng:2017usu,Sabogal:2026bsw,Feng:2019jqa,Zhang:2005yz}. More recently, Ref.~\cite{Li:2026xaz} investigated whether dark-sector interactions can reproduce the late-time expansion behavior of dynamical dark energy by considering coupled quintessence and a phenomenological coupled-fluid model\footnote{In the coupled quintessence scenario, dark energy is described by a scalar field coupled to dark matter through an interaction specified at the Lagrangian level, whereas in the coupled-fluid scenario, an interaction between dark matter and dark energy with a constant EoS parameter $w$ is introduced phenomenologically at the fluid level.}. Combining CMB, DESI DR2 BAO, and SN observations, that study found a $3$--$5\sigma$ preference for nonzero dark-sector interactions in both scenarios. Both models provide fits to current observations that are comparable to or better than those obtained with the $w_0w_a$CDM framework, suggesting that dark-sector interactions can account for the observed deviations from $\Lambda$CDM without necessarily requiring an explicitly time-dependent intrinsic dark energy EoS.

In phenomenological IDE models, however, the interaction term $Q$ is not uniquely determined from first principles, and different prescriptions can lead to distinct background and perturbation evolutions. Our previous study of IDE models with $w=-1$ showed that the observational preference for a nonzero dark-sector interaction depends sensitively on the form of $Q$, with models featuring $Q\propto\rho_{\rm de}$ generally preferred over those with $Q\propto\rho_{\rm c}$~\cite{Li:2024qso,Li:2025muv}. A natural question, therefore, is how different interaction forms affect the cosmological constraints and the statistical preference for a nonzero dark-sector interaction when the dark energy EoS is generalized from $w=-1$ to an arbitrary constant value.

In this work, we constrain IDE models with a constant dark energy EoS parameter $w$ by considering four representative forms of the interaction term $Q$, using the DESI DR2 BAO measurements, SN observations, and CMB data from Planck, Atacama Cosmology Telescope (ACT), and South Pole Telescope (SPT). Our aim is to investigate whether current cosmological observations favor a nonzero interaction between dark energy and dark matter and to determine which IDE model with a constant dark energy EoS parameter $w$ receives the strongest observational support.

This work is organized as follows. In Sec.~\ref{sec2}, we briefly introduce the IDE models and the cosmological datasets used in this work. In Sec.~\ref{sec3}, we present the parameter constraints and provide related discussion. The main conclusions are summarized in Sec.~\ref{sec4}.

\section{Methodology and Data}\label{sec2}

\subsection{Brief description of the IDE models}\label{sec2.1}

In a spatially flat Friedmann--Robertson--Walker universe, the background evolution is governed by the Friedmann equation
\begin{equation}\label{2.1}
3M_{\rm pl}^2 H^2 = \rho_{\rm de} + \rho_{\rm c} + \rho_{\rm b} + \rho_{\rm r},
\end{equation}
where $M_{\rm pl}$ is the reduced Planck mass, and $\rho_{\rm de}$, $\rho_{\rm c}$, $\rho_{\rm b}$, and $\rho_{\rm r}$ denote the energy densities of dark energy, cold dark matter (CDM), baryons, and radiation, respectively.

In IDE scenarios, we assume a direct non-gravitational energy exchange between the dark energy and CDM sectors, while baryons and radiation remain separately conserved. Consequently, the continuity equations for the dark sector are modified as follows:
\begin{gather}\label{conservation1}
\dot{\rho}_{\rm de} + 3H(1+w)\rho_{\rm de} = Q, \\
\dot{\rho}_{\rm c} + 3H\rho_{\rm c} = -Q,
\end{gather}
where the overdot denotes a derivative with respect to cosmic time $t$, $H$ is the Hubble parameter, $w$ is the constant dark energy EoS parameter, and $Q$ characterizes the rate of energy transfer between dark energy and CDM. A positive $Q$ corresponds to energy transfer from CDM to dark energy, while a negative $Q$ indicates the opposite direction.

To describe the effect of the dark-sector interaction in terms of an effective EoS, the continuity equations can be recast into an equivalent noninteracting form as
\begin{gather}
\dot{\rho}_{\rm de} + 3H(1+w_{\rm de}^{\rm eff})\rho_{\rm de} = 0, \\
\dot{\rho}_{\rm c} + 3H(1+w_{\rm c}^{\rm eff})\rho_{\rm c} = 0,
\end{gather}
where the effective EoS parameters of dark energy and CDM are given by
\begin{equation}\label{eq:w_eff}
w_{\rm de}^{\rm eff}
= w-\frac{Q}{3H\rho_{\rm de}},
\qquad
w_{\rm c}^{\rm eff}
= \frac{Q}{3H\rho_{\rm c}}.
\end{equation}
Therefore, the effective EoS of dark energy is determined jointly by its intrinsic EoS parameter $w$ and the interaction term $Q$. Even for a constant intrinsic EoS, the dark-sector interaction can induce a time-varying $w_{\rm de}^{\rm eff}$, thereby mimicking the behavior of dynamical dark energy.

We phenomenologically parameterize the interaction term $Q$, assuming that the energy transfer rate is proportional to the energy density of either dark energy or cold dark matter~\cite{Amendola:1999qq,Billyard:2000bh}. To ensure dimensional consistency, the interaction term must include a factor with dimensions of inverse time. A natural choice is the Hubble parameter $H$. On the other hand, some studies have argued that the energy exchange between dark energy and CDM should be regarded as a local process and thus should not be explicitly linked to the global expansion of the universe~\cite{Boehmer:2008av,Valiviita:2008iv,Caldera-Cabral:2008yyo,He:2008si,Clemson:2011an}. Following this consideration, the Hubble constant $H_0$ can instead be introduced solely to ensure dimensional consistency. Hence, we focus on IDE models with a constant dark energy EoS parameter $w$, hereafter referred to as I$w$CDM models. We consider four representative forms of the interaction term, namely $Q=\beta H\rho_{\rm de}$ (I$w$CDM1), $Q=\beta H\rho_{\rm c}$ (I$w$CDM2), $Q=\beta H_0\rho_{\rm de}$ (I$w$CDM3), and $Q=\beta H_0\rho_{\rm c}$ (I$w$CDM4). The dimensionless coupling parameter $\beta$ determines the strength and direction of the energy exchange: for $\beta > 0$, energy is transferred from cold dark matter to dark energy; for $\beta < 0$, the transfer is reversed; and for $\beta = 0$, no energy transfer occurs.

In studies of IDE, a long-standing issue concerns the stability of cosmological perturbations~\cite{Majerotto:2009zz,Clemson:2011an}. At the linear perturbation level, coupling a non-adiabatic dark energy fluid ($w \neq -1$) to dark matter can significantly alter the evolution of non-adiabatic pressure perturbations. Within the standard linear perturbation framework, such an interaction can induce an exponential growth of curvature perturbations on super-horizon scales, giving rise to the well-known large-scale instability problem. To self-consistently calculate the cosmological perturbations across the entire I$w$CDM parameter space, we adopt the extended parameterized post-Friedmann (ePPF) framework~\cite{Fang:2008sn,Hu:2008zd,Li:2014eha,Li:2014cee,Li:2015vla,Li:2023fdk,Zhang:2017ize,Feng:2018yew}.

\begin{figure*}[!htp]
\centering
\includegraphics[width=1\textwidth]{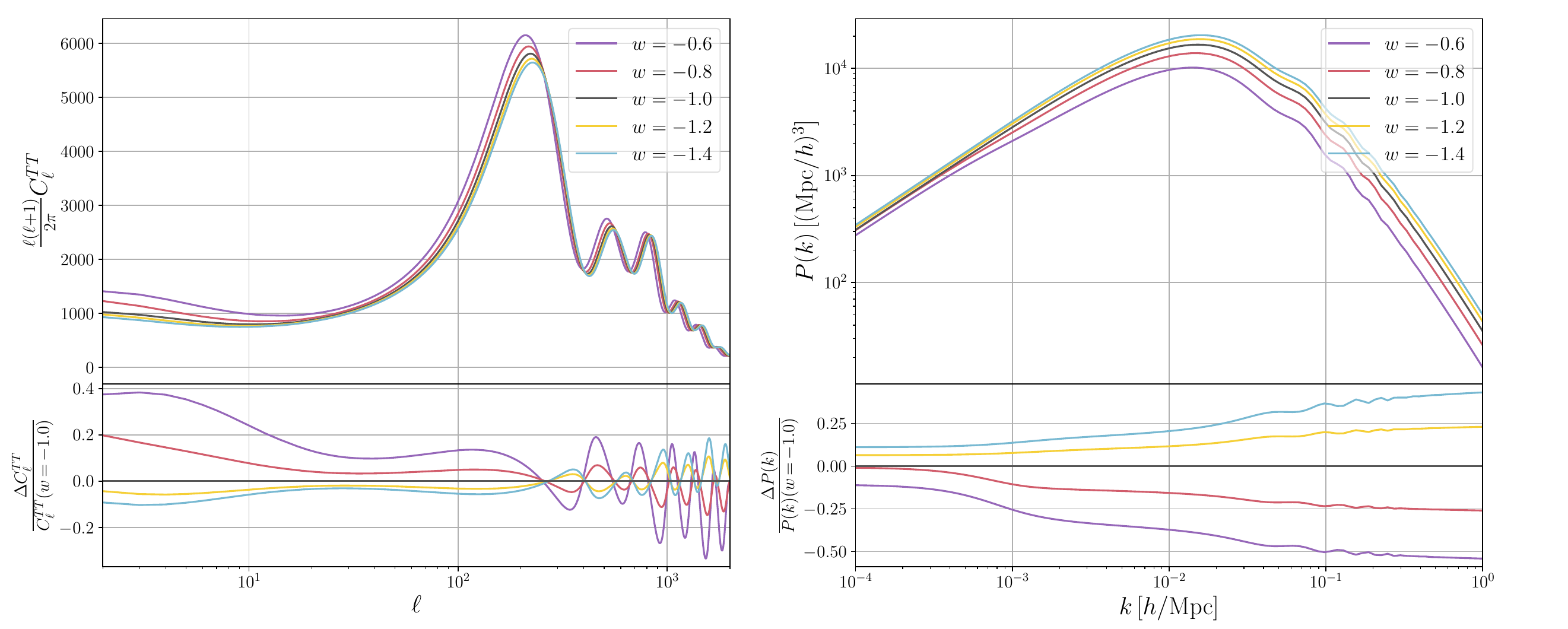}
\caption{\label{fig1}
The theoretical CMB TT power spectra (left) and matter power spectra (right) for different values of the dark energy EoS parameter $w$, with all other cosmological parameters fixed at the mean values of the IDE model reported in Ref.~\cite{Li:2024qso}.}
\end{figure*}

Figure~\ref{fig1} shows the effects of the dark energy EoS parameter $w$ on the CMB temperature angular power spectrum $C_\ell^{TT}$ (left panel) and the matter power spectrum $P(k)$ (right panel) in the I$w$CDM framework. In standard non-interacting models, dark energy remains subdominant at early times, so varying $w$ mainly modifies the late-time expansion history and has only a limited direct impact on the early-time physics that determines the primary CMB anisotropies. In the interacting case considered here, however, the situation is different. Since the dark matter evolution is coupled to the dark energy sector, changing $w$ also affects the evolution and clustering of dark matter, and consequently leaves visible signatures in both the CMB and matter power spectra.

For the CMB temperature spectrum, the main changes appear at large angular scales and around the acoustic features. Models with $w>-1$ show enhanced power at low multipoles, while those with $w<-1$ show suppressed power, reflecting changes in the late-time evolution and the corresponding integrated Sachs--Wolfe contribution. Around the acoustic region, varying $w$ changes both the peak amplitudes and their positions, with the main peaks shifting slightly toward lower $\ell$ for larger $w$ and toward higher $\ell$ for smaller $w$. The resulting differences are generally at the several-percent level around the acoustic peaks, with more pronounced oscillatory residuals appearing near the peak and trough regions at high multipoles.

For the matter power spectrum, $w>-1$ suppresses the power relative to the $w=-1$ case, whereas $w<-1$ enhances it. The modification is clearly scale dependent and becomes more pronounced toward smaller scales. For the most extreme values considered here, the deviation grows from the $\mathcal{O}(10\%)$ level on very large scales to several tens of percent at $k\gtrsim0.1\,h\,{\rm Mpc}^{-1}$, where $h$ is the dimensionless Hubble parameter. This behavior reflects the fact that, in the interacting scenario, the dark energy EoS affects the growth of matter perturbations through its coupling to the dark matter sector.

\subsection{Cosmological data}\label{sec2.2}

\begin{table}[t]
\caption{Flat priors on the main cosmological parameters constrained in this work.}
\begin{center}
\renewcommand{\arraystretch}{1.5}
\begin{tabular}{c@{\hspace{0.4cm}}@{\hspace{0.4cm}} c @{\hspace{0.4cm}} c }
\hline\hline
\textbf{Model}       & \textbf{Parameter}       & \textbf{Prior}\\
\hline
$\Lambda$CDM        & $\Omega_{\rm b} h^2$                     & $\mathcal{U}$[0.005\,,\,0.1] \\
                    & $\Omega_{\rm c} h^2$                     & $\mathcal{U}$[0.001\,,\,0.99] \\
                    & $H_0~[\mathrm{km\,s^{-1}\,Mpc^{-1}}]$    & $\mathcal{U}$[20\,,\,100] \\
                    & $\tau$                                   & $\mathcal{U}$[0.01\,,\,0.8] \\
                    & $\ln(10^{10}A_{\rm s})$                 & $\mathcal{U}$[1.61\,,\,3.91] \\
                    & $n_{\rm s}$                                  & $\mathcal{U}$[0.8\,,\, 1.2] \\
\hline
I$w$CDM             & $w$                                    & $\mathcal{U}$[-3\,,\, 1] \\
                    & $\beta$                                & $\mathcal{U}$[-5\,,\, 1] \\
			
\hline\hline
\end{tabular}
\label{tab1}
\end{center}	
\end{table}

\begin{table*}[htbp]
\caption{\label{tab2} Constraints on the cosmological parameters in the $\Lambda$CDM and four I$w$CDM models using the CMB, DESI, and SN data.}
\renewcommand{\arraystretch}{1.35}
\setlength{\tabcolsep}{3pt}
\scriptsize
\begin{ruledtabular}
\begin{tabular*}{\textwidth}{@{\extracolsep{\fill}}lcccccc}
\begin{tabular}{l}\textbf{Model / }\textbf{Dataset}\end{tabular} &
\begin{tabular}{c}\textbf{$H_0$}~\textbf{$[\mathrm{km\,s^{-1}\,Mpc^{-1}}]$}\end{tabular} &
\textbf{$\Omega_\mathrm{m}$} & \textbf{$\sigma_8$} & \textbf{$S_8$} & \textbf{$\beta$} & \textbf{$w$} \\
\hline
\multicolumn{7}{l}{\textbf{$\Lambda$CDM}} \\
CMB+DESI & $68.20 \pm 0.26$ & $0.3031^{+0.0033}_{-0.0037}$ & $0.8156 \pm 0.0048$ & $0.8199 \pm 0.0067$ & -- & -- \\
CMB+DESI+DES-Dovekie & $68.10 \pm 0.25$ & $0.3046 \pm 0.0034$ & $0.8153^{+0.0044}_{-0.0050}$ & $0.8215 \pm 0.0064$ & -- & -- \\
CMB+DESI+DESY5 & $68.04 \pm 0.25$ & $0.3054 \pm 0.0034$ & $0.8153^{+0.0046}_{-0.0052}$ & $0.8226 \pm 0.0067$ & -- & -- \\
CMB+DESI+PantheonPlus & $68.10 \pm 0.26$ & $0.3046 \pm 0.0035$ & $0.8154^{+0.0045}_{-0.0050}$ & $0.8216 \pm 0.0067$ & -- & -- \\
\hline
\multicolumn{7}{l}{\textbf{I$w$CDM1}} \\
CMB+DESI & $69.45 \pm 0.88$ & $0.5420^{+0.1200}_{-0.0820}$ & $0.5300^{+0.0340}_{-0.0910}$ & $0.7000^{+0.0170}_{-0.0370}$ & $-1.92^{+1.20}_{-0.92}$ & $-1.690^{+0.420}_{-0.290}$ \\
CMB+DESI+DES-Dovekie & $67.77 \pm 0.51$ & $0.6190^{+0.1000}_{-0.0840}$ & $0.4860^{+0.0370}_{-0.0620}$ & $0.6910^{+0.0140}_{-0.0230}$ & $-2.70^{+1.10}_{-1.70}$ & $-1.870 \pm 0.400$ \\
CMB+DESI+DESY5 & $67.39 \pm 0.52$ & $0.5600^{+0.0690}_{-0.1000}$ & $0.5240^{+0.0540}_{-0.0640}$ & $0.7090 \pm 0.0220$ & $-1.84^{+1.20}_{-0.38}$ & $-1.580^{+0.390}_{-0.120}$ \\
CMB+DESI+PantheonPlus & $67.93 \pm 0.58$ & $0.5200 \pm 0.1000$ & $0.5540^{+0.0710}_{-0.1100}$ & $0.7180^{+0.0260}_{-0.0410}$ & $-1.56^{+1.20}_{-0.97}$ & $-1.510^{+0.400}_{-0.320}$ \\
\hline
\multicolumn{7}{l}{\textbf{I$w$CDM2}} \\
CMB+DESI & $69.41 \pm 0.92$ & $0.2931 \pm 0.0071$ & $0.8300 \pm 0.0100$ & $0.8205 \pm 0.0066$ & $0.0018^{+0.0014}_{-0.0016}$ & $-1.027 \pm 0.043$ \\
CMB+DESI+DES-Dovekie & $68.07 \pm 0.54$ & $0.3028 \pm 0.0050$ & $0.8178 \pm 0.0080$ & $0.8215 \pm 0.0066$ & $0.0030 \pm 0.0014$ & $-0.962 \pm 0.024$ \\
CMB+DESI+DESY5 & $67.64 \pm 0.54$ & $0.3057 \pm 0.0050$ & $0.8137^{+0.0083}_{-0.0072}$ & $0.8214 \pm 0.0067$ & $0.0035 \pm 0.0014$ & $-0.940 \pm 0.025$ \\
CMB+DESI+PantheonPlus & $68.21 \pm 0.56$ & $0.3016 \pm 0.0050$ & $0.8191 \pm 0.0078$ & $0.8213 \pm 0.0066$ & $0.0030 \pm 0.0014$ & $-0.968 \pm 0.026$ \\
\hline
\multicolumn{7}{l}{\textbf{I$w$CDM3}} \\
CMB+DESI & $67.40 \pm 1.10$ & $0.5890 \pm 0.0730$ & $0.4970^{+0.0290}_{-0.0500}$ & $0.6910^{+0.0110}_{-0.0180}$ & $-2.28^{+0.90}_{-0.81}$ & $-1.450 \pm 0.150$ \\
CMB+DESI+DES-Dovekie & $67.49 \pm 0.50$ & $0.6100 \pm 0.0520$ & $0.4810^{+0.0260}_{-0.0310}$ & $0.6830^{+0.0110}_{-0.0140}$ & $-2.52^{+0.80}_{-0.54}$ & $-1.510^{+0.170}_{-0.120}$ \\
CMB+DESI+DESY5 & $66.90 \pm 0.56$ & $0.6520^{+0.0680}_{-0.0450}$ & $0.4630^{+0.0160}_{-0.0330}$ & $0.6796^{+0.0078}_{-0.0130}$ & $-3.07 \pm 0.83$ & $-1.590 \pm 0.180$ \\
CMB+DESI+PantheonPlus & $67.62 \pm 0.57$ & $0.5970 \pm 0.0600$ & $0.4890^{+0.0280}_{-0.0410}$ & $0.6860^{+0.0120}_{-0.0180}$ & $-2.40^{+0.90}_{-0.59}$ & $-1.490^{+0.200}_{-0.120}$ \\
\hline
\multicolumn{7}{l}{\textbf{I$w$CDM4}} \\
CMB+DESI & $69.91 \pm 0.99$ & $0.3010 \pm 0.0130$ & $0.8310 \pm 0.0110$ & $0.8330 \pm 0.0150$ & $-0.0510^{+0.0490}_{-0.0550}$ & $-1.106 \pm 0.059$ \\
CMB+DESI+DES-Dovekie & $67.91 \pm 0.54$ & $0.3070^{+0.0120}_{-0.0140}$ & $0.8125 \pm 0.0078$ & $0.8220 \pm 0.0150$ & $-0.0060 \pm 0.0520$ & $-0.996 \pm 0.035$ \\
CMB+DESI+DESY5 & $67.46 \pm 0.53$ & $0.3100 \pm 0.0130$ & $0.8084 \pm 0.0077$ & $0.8220 \pm 0.0150$ & $-0.0020 \pm 0.0520$ & $-0.977^{+0.038}_{-0.032}$ \\
CMB+DESI+PantheonPlus & $68.07 \pm 0.58$ & $0.3080 \pm 0.0130$ & $0.8141 \pm 0.0080$ & $0.8250 \pm 0.0150$ & $-0.0160 \pm 0.0500$ & $-1.009^{+0.039}_{-0.034}$ \\
\end{tabular*}
\end{ruledtabular}
\end{table*}

\begin{figure*}[!t]
\centering
\includegraphics[width=0.7\textwidth]{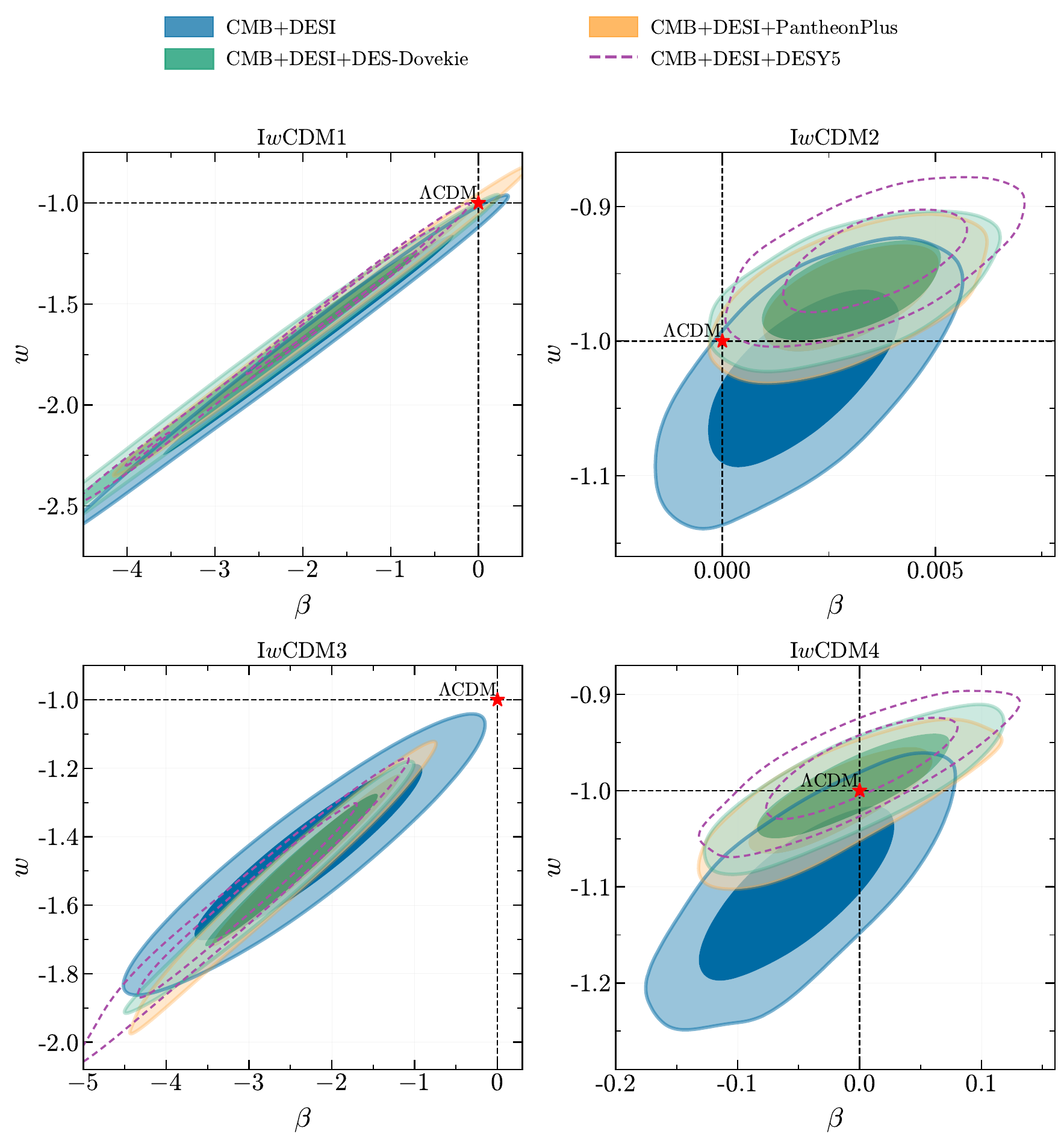}
\caption{\label{fig2}
Two-dimensional marginalized constraints in the $\beta$-$w$ plane for the four I$w$CDM models using the CMB, DESI, and SN data. The vertical and horizontal dashed lines denote $\beta=0$ and $w=-1$, respectively, while the red star marks the $\Lambda$CDM limit $(\beta,w)=(0,-1)$.}
\end{figure*}

The theoretical predictions for the background and perturbation evolution of the different models are computed using the \texttt{IDECAMB} package~\cite{Li:2023fdk}, a modified version of \texttt{CAMB}~\cite{Lewis:1999bs} specifically tailored for IDE scenarios. To self-consistently calculate cosmological perturbations in the I$w$CDM models, \texttt{IDECAMB} incorporates the ePPF framework, which accounts for the dark sector coupling and avoids the unphysical large-scale instabilities that can arise in IDE models. For parameter estimation, the calculations are integrated within the public Bayesian analysis framework \texttt{Cobaya}~\cite{Torrado:2020dgo}, and the posterior probability distributions are derived using Markov chain Monte Carlo (MCMC) sampling~\cite{Lewis:2002ah,Lewis:2013hha}. The convergence of the MCMC chains is verified using the Gelman--Rubin criterion ($\hat{R}-1<0.02$), while post-processing and visualization of the marginalized posterior distributions are performed with the \texttt{GetDist} package~\cite{Lewis:2019xzd}.

The priors adopted for the sampled cosmological parameters are listed in Table~\ref{tab1}. The datasets are detailed below:
\begin{itemize}
    \item \textbf{CMB}: For the Planck data, the large-scale temperature (TT) and E-mode polarization (EE) spectra in the range $2 \le \ell \le 30$ are evaluated using the \texttt{Commander} and \texttt{SimAll} likelihoods, respectively. On smaller scales, we adopt the \texttt{CamSpec} likelihood for the TT ($30 \le \ell \le 1000$) and TE/EE ($30 \le \ell \le 600$) spectra~\cite{Efstathiou:2019mdh,Rosenberg:2022sdy}. We additionally incorporate high-multipole measurements from the ACT DR6~\cite{AtacamaCosmologyTelescope:2025blo} and SPT-3G~\cite{SPT-3G:2025bzu} 2019--2020 Main Field observations via the \texttt{ACT-lite} and \texttt{SPT-lite} likelihoods. The ACT and SPT TT spectra cover $1000 \le \ell \le 8500$, whereas the TE/EE spectra span $600 \le \ell \le 8500$. To avoid mode double-counting, the Planck TT and TE/EE spectra are restricted to $\ell < 1000$ and $\ell < 600$, respectively, following the ACT methodology. We also include the joint CMB lensing power spectrum reconstructed from the Planck, ACT, and SPT-3G maps~\cite{ACT:2025qjh, Carron:2022eyg, ACT:2023dou, ACT:2023kun}.
\end{itemize}

\begin{itemize}
    \item \textbf{DESI}: The BAO measurements from the DESI DR2 using galaxies, quasars, and the Lyman-$\alpha$ forest are reported in Ref.~\cite{DESI:2025zgx}. These measurements span a redshift range of $0.1 < z < 4.2$ and constrain the transverse comoving distance $D_\mathrm{M}(z)/r_\mathrm{d}$, the angle-averaged distance $D_\mathrm{V}(z)/r_\mathrm{d}$, and the Hubble distance $D_\mathrm{H}(z)/r_\mathrm{d}$, with all distances normalized by the comoving sound horizon $r_\mathrm{d}$ at the drag epoch.
\end{itemize}

\begin{itemize}
    \item \textbf{PantheonPlus}: The PantheonPlus compilation contains 1550 spectroscopically confirmed type~Ia supernovae (SNe) from 18 independent surveys, covering the redshift range $0.01 < z < 2.26$~\cite{Brout:2022vxf}.
\end{itemize}

\begin{itemize}
    \item \textbf{DESY5}: The DESY5 compilation consists of 1829 SNe, including 1635 high-redshift photometrically classified events from the full 5-year Dark Energy Survey sample over $0.1 < z < 1.3$, together with a low-redshift anchor sample of 194 SNe over $0.025 < z < 0.1$~\cite{DES:2024jxu}.
\end{itemize}

\begin{itemize}
    \item \textbf{DES-Dovekie}: The DES-Dovekie dataset is a newly reconstructed SN sample based on a recalibration and re-analysis of the DESY5 data~\cite{DES:2025sig}. It contains 1623 likely SNe with $P_{\rm SNe}>0.5$ and 197 low-redshift SNe, for a total of 1820 SNe. Compared with the original DESY5 sample, DES-Dovekie features recalibrated photometric measurements and a re-analysis of the SN data.
\end{itemize}

\section{RESULTS AND DISCUSSION}\label{sec3}

\begin{figure}[!htp]
\centering
\hspace*{-1.05cm}
\includegraphics[width=1.05\columnwidth]{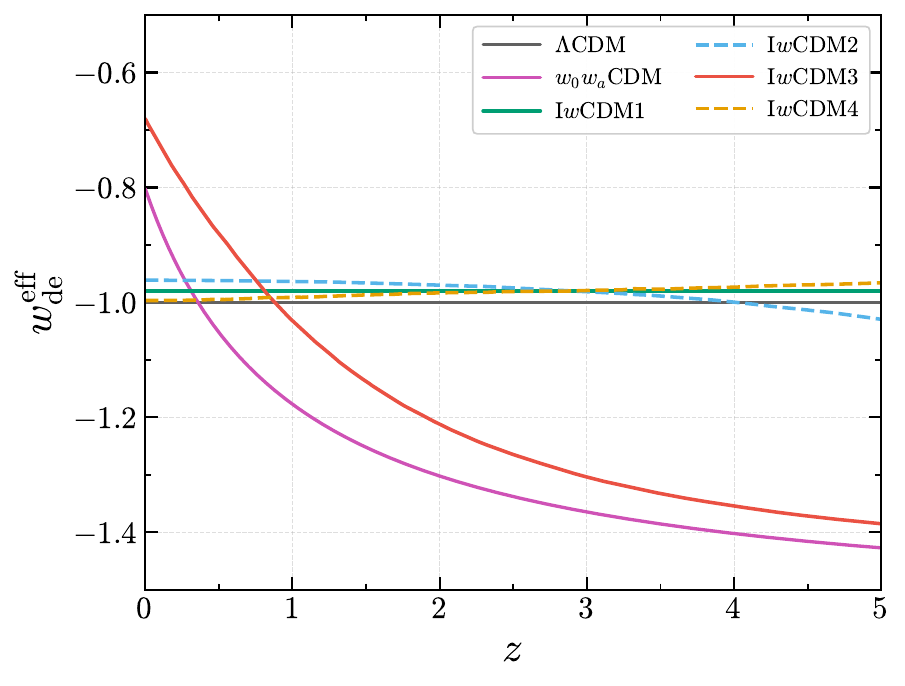}
\caption{\label{fig3}Evolution of the effective dark energy EoS, $w_{\rm de}^{\rm eff}$, for the four I$w$CDM models, compared with the $\Lambda$CDM and $w_0w_a$CDM models. The I$w$CDM curves are computed using the best-fit parameters inferred from the CMB+DESI+DES-Dovekie data.}
\end{figure}

\begin{figure*}[!htp]
\centering
\includegraphics[width=0.8\textwidth]{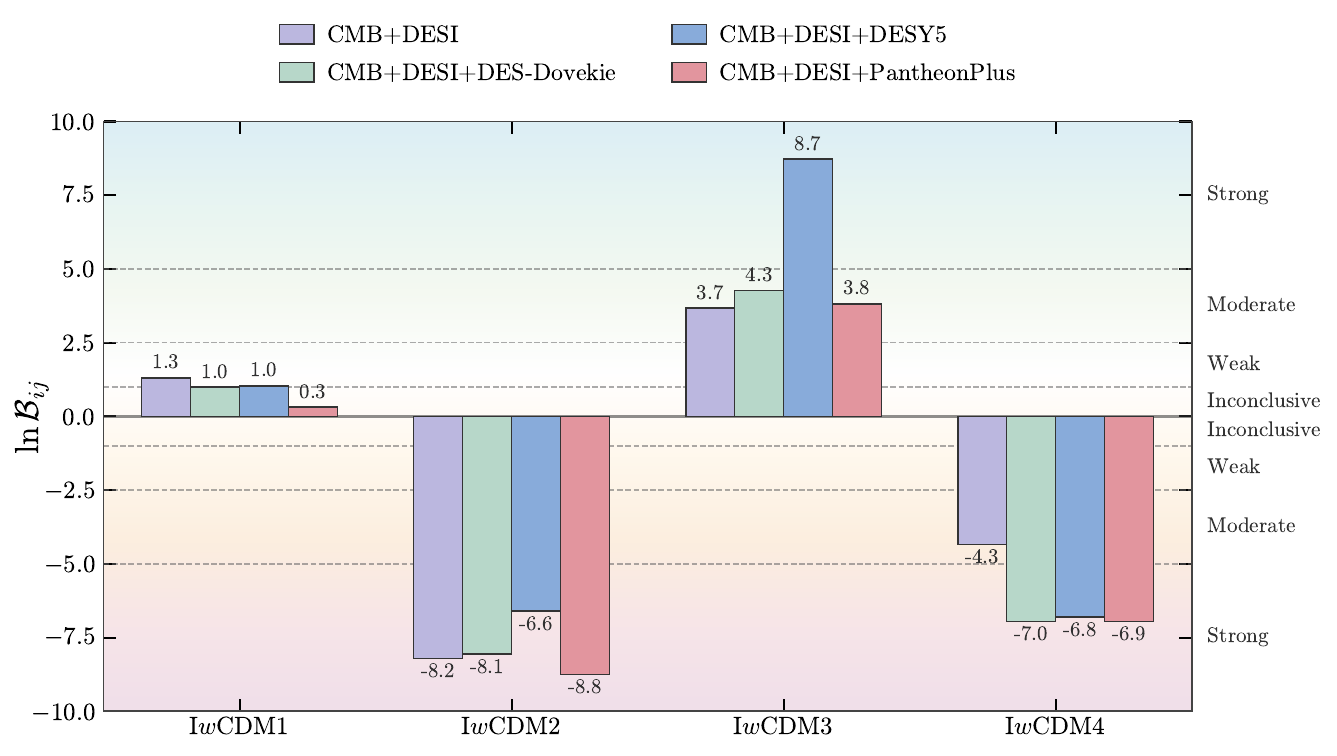}
\caption{\label{fig4}
Bayes factors ($\ln \mathcal{B}_{ij}$) for the four I$w$CDM models relative to $\Lambda$CDM. Positive values favor the respective I$w$CDM model, whereas negative values favor $\Lambda$CDM.}
\end{figure*}

We constrain the four I$w$CDM models using the latest datasets, combining the CMB, DESI BAO, and SN samples from DES-Dovekie, DESY5, and PantheonPlus. The resulting constraints on the cosmological parameters are summarized in Table~\ref{tab2}. The two-dimensional marginalized contours of $\beta$ and $w$ for the four models are shown in Fig.~\ref{fig2}. We then present the redshift evolution of the effective dark energy EoS in Fig.~\ref{fig3}. Finally, we compare the Bayes factors of the different models in Fig.~\ref{fig4}.

As shown in Table~\ref{tab2} and Fig.~\ref{fig2}, among the models with an interaction rate proportional to the dark energy density, the I$w$CDM3 model generally shows more pronounced departures from the $\Lambda$CDM limit than the I$w$CDM1 model. Using CMB+DESI+DES-Dovekie, the I$w$CDM3 model gives $\beta=-2.52^{+0.80}_{-0.54}$ and $w=-1.510^{+0.170}_{-0.120}$, corresponding to deviations from $\beta=0$ and $w=-1$ at approximately the $3.2\sigma$ and $3.0\sigma$ levels, respectively. The significance of these departures depends on the SN compilation. DESY5 yields the strongest preference, whereas PantheonPlus yields the weakest. The DES-Dovekie sample was constructed through a recalibration and reanalysis of DESY5 to address calibration inconsistencies between its low- and high-redshift subsamples. This trend is consistent with the DESI results for dynamical dark energy, where the evidence for deviations from $\Lambda$CDM also depends sensitively on the adopted SN sample.

It is worth noting that a distinctive feature of the I$w$CDM1 and I$w$CDM3 models is that they systematically favor a higher matter abundance accompanied by a strongly suppressed clustering amplitude across all dataset combinations. For example, for the CMB+DESI+DES-Dovekie data, I$w$CDM1 yields $\Omega_{\rm m}=0.619^{+0.100}_{-0.084}$ and $\sigma_8=0.486^{+0.037}_{-0.062}$, while I$w$CDM3 gives $\Omega_{\rm m}=0.610\pm0.052$ and $\sigma_8=0.481^{+0.026}_{-0.031}$. This behavior arises because $\beta<0$ in these models corresponds to a continuous transfer of energy from dark energy to CDM, causing the CDM energy density to dilute more slowly with cosmic expansion than in $\Lambda$CDM and consequently leading to a substantially enhanced present-day matter density. At the background level, the increased matter density is compensated by the phantom-like dark energy EoS ($w<-1$) favored by the fit, allowing the models to remain consistent with background distance measurements despite the unusually large present-day matter density. At the perturbation level, however, these models exhibit a strongly suppressed growth of structure. The energy exchange between dark energy and CDM, together with the phantom-like EoS, substantially enhances the Hubble friction and thereby suppresses structure formation at late times. Consequently, despite the significantly enhanced present-day matter abundance, the linear growth factor is reduced, driving $\sigma_8$ down to approximately $0.5$.

For the I$w$CDM2 and I$w$CDM4 models, since CDM accounts for a much larger fraction of the cosmic energy budget than dark energy in the early universe, the interaction in these two models can significantly affect the background and perturbation evolution during the epochs probed by the CMB. The coupling parameter is therefore tightly constrained, as any sizable deviation from $\beta=0$ would induce changes in the CMB observables beyond the range allowed by the data. For example, using the CMB+DESI+DES-Dovekie data, we obtain $\beta=0.0030\pm0.0014$ and $w=-0.962\pm0.024$ in I$w$CDM2, compared to $\beta=-0.006\pm0.052$ and $w=-0.996\pm0.035$ in I$w$CDM4. The former corresponds to an approximately $2.1\sigma$ preference for a nonzero coupling, although the central value of $\beta$ remains very close to zero and hence indicates only a weak interaction. The difference between these two models can be traced to the Hubble factor entering the interaction term. Since $H$ is much larger than $H_0$ in the early universe, the interaction in I$w$CDM2 is more strongly amplified at early times and is therefore subject to a much tighter CMB constraint, forcing $\beta$ into a very narrow region around zero.

Figure~\ref{fig3} shows the evolution of the effective dark energy EoS, $w_{\rm de}^{\rm eff}$ in Eq.~\eqref{eq:w_eff}, reconstructed from the best-fit parameters for the CMB+DESI+DES-Dovekie data. By incorporating the energy transfer between CDM and dark energy into the evolution of the dark energy density, $w_{\rm de}^{\rm eff}$ provides a useful description of the background dynamics induced by the interaction, even when the intrinsic dark energy EoS is constant. The most distinctive behavior is found in the I$w$CDM3 model. Its effective EoS evolves from the phantom regime at high redshift, crosses the phantom divide $w_{\rm de}^{\rm eff}=-1$, and enters the quintessence-like region toward low redshift. The resulting evolution closely resembles the behavior generated by the $w_0w_a$CDM model. Therefore, although I$w$CDM3 contains no explicit time dependence in the intrinsic dark energy EoS, the energy exchange in the dark sector allows it to reproduce an effective dynamical-dark energy behavior at the background level. This pronounced redshift evolution of the effective dark energy EoS allows I$w$CDM3 to better accommodate the deviations indicated by the BAO and SN data, and I$w$CDM3 is therefore expected to provide a significantly improved fit relative to $\Lambda$CDM. Such behavior is absent in the other IDE models. The effective EoS of I$w$CDM1 remains nearly constant, while the much smaller couplings allowed in I$w$CDM2 and I$w$CDM4 prevent them from exhibiting a redshift evolution as pronounced as that of the $w_0w_a$CDM model over the low- and intermediate-redshift range.

To assess the statistical preference for the different I$w$CDM models relative to $\Lambda$CDM, we employ Bayesian evidence as the model selection criterion, computed directly from the MCMC chains using the {\tt MCEvidence} package~\cite{Heavens:2017hkr,Heavens:2017afc}. Model comparison is performed using the logarithmic Bayes factor, defined as $\ln \mathcal{B}_{ij}=\ln Z_i-\ln Z_j$, where $Z_i$ and $Z_j$ denote the Bayesian evidences for models $i$ and $j$, respectively. The strength of the evidence is interpreted according to the revised Jeffreys scale~\cite{Kass:1995loi,Trotta:2008qt}.

As shown in Fig.~\ref{fig4}, we compare the four I$w$CDM models using the logarithmic Bayes factor $\ln\mathcal{B}_{ij}$, where $i$ and $j$ denote the I$w$CDM and $\Lambda$CDM models, respectively, and positive values favor I$w$CDM. We find that I$w$CDM1 and I$w$CDM3 are favored over $\Lambda$CDM to varying degrees, whereas I$w$CDM2 and I$w$CDM4 are substantially disfavored. This indicates that models with $Q\propto\rho_{\rm de}$ generally receive stronger observational support than those with $Q\propto\rho_{\rm c}$. In particular, for the CMB+DESI+DES-Dovekie data combination, I$w$CDM3 yields $\ln\mathcal{B}_{ij}=4.3$, indicating moderate Bayesian evidence in its favor over $\Lambda$CDM. This result is consistent with the effective EoS evolution shown in Fig.~\ref{fig3}, where I$w$CDM3 exhibits a redshift evolution as pronounced as that of the $w_0w_a$CDM model, allowing it to better accommodate the DESI BAO and SN observations. Overall, these findings highlight the strong dependence of the observational preference for IDE on the assumed interaction form, with I$w$CDM3 receiving the strongest statistical support among the four models considered.

\section{Conclusion}\label{sec4}

In this work, we aim to investigate whether dark energy and dark matter interact and which interaction forms are favored. We constrain four representative IDE models with a constant dark energy EoS parameter $w$ using current cosmological data.

We find that the IDE models with $Q\propto\rho_{\rm de}$ generally show a stronger preference for a nonzero dark-sector interaction than those with $Q\propto\rho_{\rm c}$. For the CMB+DESI+DES-Dovekie data, I$w$CDM1 yields $\beta=-2.70^{+1.10}_{-1.70}$ and $w=-1.870\pm0.400$, while I$w$CDM3 gives $\beta=-2.52^{+0.80}_{-0.54}$ and $w=-1.510^{+0.170}_{-0.120}$, with the latter showing deviations from $\beta=0$ and $w=-1$ at approximately the $3.2\sigma$ and $3.0\sigma$ levels, respectively. We also find that the effective EoS of I$w$CDM3 exhibits a pronounced redshift evolution, crossing the phantom divide and closely resembling the behavior of the $w_0w_a$CDM model. By comparison, the effective EoS of I$w$CDM1 remains nearly constant, while I$w$CDM2 and I$w$CDM4 show much weaker evolution. The Bayesian evidence further indicates that models with $Q\propto\rho_{\rm de}$ are more favored by current observations, particularly I$w$CDM3, which yields $\ln\mathcal{B}_{ij}=4.3$, corresponding to moderate evidence in its favor over $\Lambda$CDM.

Overall, I$w$CDM3 shows both the strongest preference for a nonzero dark-sector interaction and the strongest observational support among the four models considered. Future precision late-time observations from the Large Synoptic Survey Telescope, Euclid, and subsequent DESI data releases will help determine whether the preference for dark-sector interactions persists and to distinguish between different interaction forms.

\section*{Acknowledgments}
This work was supported by the National Natural Science Foundation of China (Grants Nos. 12533001, 12575049, and 12473001), the National SKA Program of China (Grants Nos. 2022SKA0110200 and 2022SKA0110203), the China Manned Space Program (Grant No. CMS-CSST-2025-A02), and the National 111 Project (Grant No. B16009).

\bibliography{main}

\end{document}